# Near IR bandgap semiconductive 2D conjugated metal-organic framework with rhombic lattice and high mobility

Lukas Sporrer,[#][a] Guojun Zhou,[#][b] Mingchao Wang,[#][a] Vasileios Balos,[c] Sergio Revuelta,[c] Kamil Jastrzembski,[a] Markus Löffler,[d] Petko Petkov,[e] Thomas Heine,[a] Angieszka Kuc,*[f] Enrique Cánovas,*[c] Zhehao Huang,*[b] Xinliang Feng,*[a,g] and Renhao Dong*[a,h]

[a] Mr. L. Sporrer, Dr. M. Wang, Mr. K. Jastrzembski, Prof. T. Heine, Prof. Dr. X. Feng, Prof. Dr. R. Dong
Chair of Molecular Functional Materials, Faculty of Chemsitry and Food Chemistry
Technische Universität Dresden, Mommsenstrasse 4, Dresden 01062, Germany
E-mail: xinliang.feng@tu-dresden.de

[b] Dr. G. Zhou, Dr. Z. Huang
Department of Materials and Environmental Chemistry, Stockholm University, Stockholm SE-106 91, Sweden
Email: zhehao.huang@mmk.su.se

[c] Mr. V. Balos, Mr. S. Revuelta, Prof. Dr. E. Cánovas
Instituto Madrileño de Estudios Avanzados en Nanociencia (IMDEA Nanociencia), Madrid 28049, Spain
Email: enrique.canovas@imdea.org

[d] Dr. M. Löffler
Dresden Centre for Nanoanalysis, Technische Universität Dresden, Helmholtzstr. 18, Dresden 01062, Germany

[e] Prof. Dr. P. Petkov
University of Sofia, Faculty of Chemistry and Pharmacy, 1164 Sofia (Bulgaria)

[f] Dr. A. Kuc
Helmholtz-Zentrum Dresden-Rossendorf, Abteilung Ressourcenökologie, Forschungsstelle Leipzig, 04318 Leipzig, Germany
Email: a.kuc@hzdr.de

[g] Prof. Dr. X. Feng
Max-Planck-Institut für Mikrostrukturphysik, Weinberg 2, D-06120 Halle

[h] Prof. Dr. R. Dong
Key Laboratory of Colloid and Interface Chemistry of the Ministry of Education, School of Chemistry and Chemical Engineering, Shandong University, Jinan, 250100, China
Email: renhaodong@sdu.edu.cn

# L.S., G.Z. and M.W. contributed equally to this work.

**Abstract:** Two-dimensional conjugated metal-organic frameworks (2D *c*-MOFs) are emerging as a unique class of 2D electronic materials. However, intrinsically semiconducting 2D *c*-MOFs with gaps in the Vis-NIR and high charge carrier mobility have been rare. Most of the reported semiconducting 2D *c*-MOFs are metallic (i.e. gapless), which limits their use in applications where larger band gaps are needed for logic devices. Herein, we design a new $D_{2h}$-geometric ligand, 2,3,6,7,11,12,15,16 – octahydroxyphenanthro-[9,10:b]triphenylene (OHPTP), and synthesize the first example of a 2D *c*-MOF single crystal (OHPTP-Cu) with a rhombohedral pore geometry after coordination with copper. The continuous rotation electron diffraction (cRED) analysis unveils the orthorhombic crystal structure at the atomic level with a unique AB layer stacking. The resultant $Cu_2$(OHPTP) is a *p*-type semiconductor with an indirect band gap of ~0.50 eV and exhibits high electrical conductivity of 0.10 S cm$^{-1}$ and high charge carrier mobility of 10.0 cm$^2$V$^{-1}$s$^{-1}$. Density-functional theory calculations underline the predominant role of the out-of-plane charge transport in this semiquinone-based 2D *c*-MOFs.

Introduction

Metal-organic frameworks (MOFs) are a versatile class of coordination polymers with high crystallinity, large surface areas and pore volumes that have attracted considerable attention in a wide variety of fields, such as gas storage and separation, sensing and catalysis.[1] However, most of the three-dimensional (3D) MOFs developed to date are electrical insulators, which limits their applicability in areas that require long range charge transport, as those in the fields of electronics and energy storage.[2] To fill this gap, two-dimensional conjugated MOFs (2D *c*-MOFs), a class of layer-stacked MOFs with in-plane extended π-conjugation and out-of-plane π-π stacking interactions, have exhibited uniquely record-high electrical conductivities (up to $10^3$ S cm$^{-1}$).[3] In 2012, the first example of a 2D *c*-MOF was reported based on the $C_3$ symmetric ligand hexahydroxytriphenylene (HHTP).[4] Since then, various planar, $C_6$ (e.g. benzene, coronene)[5], $C_3$ (e.g. triphenylene)[6] or $C_4$ symmetric (e.g. phthalocyanine)[7] conjugated ligands containing ortho-substituted linking groups (e.g. OH, NH$_2$, SH, SeH) have been developed. Owing to the abundant, well-defined metal-heteroatom coordination sites, high charge transport properties, redox activity and porosity, the novel subclass of 2D *c*-MOFs have attracted substantial interest in opto-electronics,[8] spintronics,[9] and energy storage devices.[10]

While being good electrical conductors, so far, most of the reported 2D *c*-MOFs reveal a band gapless electronic structure.[11] Till now, only few examples based on semiquinone-based 2D *c*-MOFs with honeycomb lattice have been discovered to be intrinsic semiconductors, all of them with relatively low charge-carrier mobilities of below 1 cm$^2$V$^{-1}$s$^{-1}$ for bulk crystals and 5 cm$^2$V$^{-1}$s$^{-1}$ for nanosheets.[12] This aspect limits the applicability of 2D MOFs in optoelectronics, where an appreciable finite band gap in needed. Therefore, it is highly attractive to develop novel 2D *c*-MOF semiconductors with wider band gaps having large carrier mobilities and high porosities.[13]

Herein, we address this challenge by developing a planar, conjugated ligand, 2,3,6,7,11,12,15,16- octahydroxyphenanthro [9,10:b]triphenylene (OHPTP), which has a $D_{2h}$ symmetry and four connecting points. From this unique π-extended ligand, we have further prepared an OHPTP-based CuO$_4$-linked 2D *c*-MOF, Cu$_2$(OHPTP), which is the first example of a rhombic single-crystalline 2D *c*-MOF. Notably, the porous samples (surface area up to 741 m$^2$ g$^{-1}$) can be obtained as single crystals with rod-like morphology with lengths in the μm range and widths of several hundred nanometers. The rhombic crystal structure of Cu$_2$(OHPTP) was resolved using continuous rotation electron diffraction (cRED) at the atomic level, providing an alternating AB stacking sequence in which two adjacent layers are shifted along one direction. Density-functional theory (DFT) calculations of the electronic structures of Cu$_2$(OHPTP) reveal that the samples are semiconducting with an indirect band gap of ~0.50 eV. The band dispersion is more pronounced in the out-of-plane direction. This Cu$_2$(OHPTP) 2D *c*-MOF displays a relatively high bulk DC conductivity of 0.10 S cm$^{-1}$ at room temperature; estimated by the 4-probe method. This figure is validated in parallel from THz spectroscopy, which was also employed to provide information about the conductivity in the AC limit. From modelling, we obtain a high charge carrier mobility of ~10.0 cm$^2$V$^{-1}$s$^{-1}$ for a doping content of ~5·10$^{16}$ cm$^{-3}$ under ambient conditions, a mobility which is primarily limited by grain boundary scattering (i.e. that will be dramatically boosted to several hundreds of cm$^2$V$^{-1}$s$^{-1}$ in larger crystals). These charge transport figures, conductivity and mobility, of merit are far superior when compared to those reported for semiquinone-based 2D *c*-MOFs.[14]

Results and Discussion

2,3,6,7,11,12,15,16-octahydroxyphenanthro[9,10:b]triphenylene (OHPTP) was prepared through a three-step synthesis with an overall yield of 79 %. After a fourfold Suzuki coupling of 1,2,4,5-tetrabromobenzene with 3,4-dimethoxyphenylboronic acid, the phenanthrotriphenylene (PTP) core was then constructed by a FeCl$_3$-promoted Scholl reaction. Finally, BBr$_3$ was employed to deprotect the eight hydroxyl groups on PTP yielding OHPTP as a light green solid (Figure 1a, Scheme S1, detailed synthetic conditions shown in SI).

The 2D *c*-MOF synthesis was performed by coordination of the OHPTP ligand with Cu(II) under solvothermal conditions (Figure 1a). After optimization of copper source, solvents and temperature (Scheme S2, Figure S1), Cu$_2$(OHPTP) was prepared in a neutral water-DMF (7:3) mixture at 85 °C for 24 h (Figure 1a) to yield a black powder. Scanning electron microscopy (SEM) indicated that the bulk sample consisted of rather homogeneous rod-like crystals with lengths in the μm range and widths of several hundreds of nanometers (Figure 1b). Energy dispersive X-ray spectroscopy (EDX) showed a uniform distribution of copper, carbon and oxygen throughout the single crystals.

Fourier-transform infrared spectroscopy (FT-IR) showed the disappearance of the characteristic OH stretching vibration of the OHPTP ligands at around 3500 - 3000 cm$^{-1}$, the attenuation of the C-OH stretching vibration at 1021 cm$^{-1}$ as well as the attenuation of the C=O stretching vibration at 1523 cm$^{-1}$, which demonstrated the formation of semiquinone form after the coordination (Figure S2a). Raman spectroscopy revealed the appearance of two characteristic bands with a broad one at 1350 to 1400 cm$^{-1}$ and another at 1575 cm$^{-1}$ (Figure S2b). These Raman modes are similar to the D and G bands found in graphene or graphite samples.[15] X-ray photoemission spectroscopy (XPS) confirmed the presence of copper, oxygen and carbon in the sample (Figure S3). High-resolution XPS of the O 2p region shows two oxygen peaks at 531.7 and 533.2 eV, corresponding to the ligand in the semiquinone form.[16] This is further confirmed by the carbon spectrum which features three distinct peaks at 284.7, 286.2 and 288.6 eV corresponding to C=C, *C*-O and C=O carbon atoms.[17] The high-resolution XPS analysis of Cu 2p revealed the dominant presence of Cu(II) with a binding energy of 934,6 eV. The combination of elemental analysis (EA) and inductively coupled plasma optical emission spectroscopy (ICP-OES) indicated the Cu/C ratio as 0.37, which agrees well with the expected 0.35 for the proposed structure (Table S1).

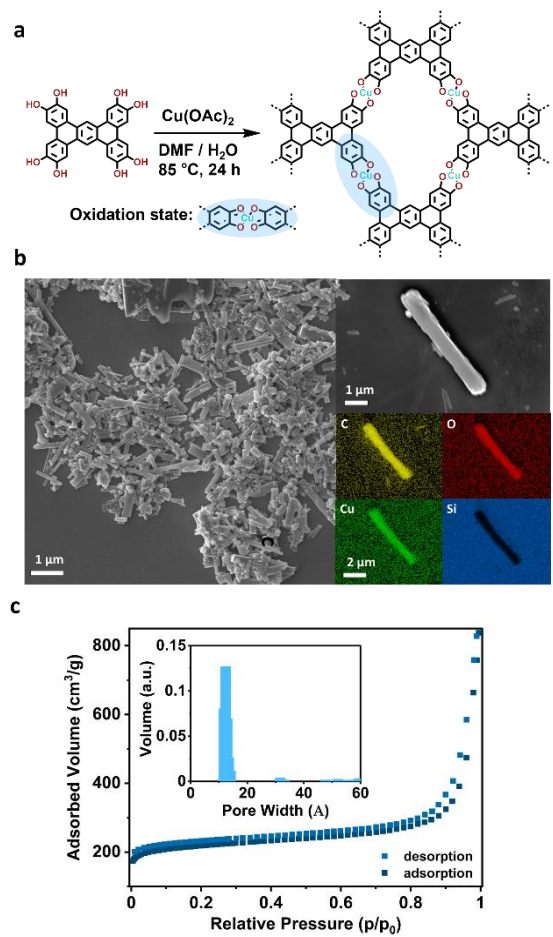

**Figure 1.** (a) Synthesis scheme of Cu$_2$(OHPTP). (b) SEM of Cu$_2$(OHPTP) crystals and EDX of a single crystal. (c) N$_2$ adsorption measurements at 77 K.

Nitrogen adsorption measurements at 77 K suggested the nano porous nature of this coordination polymer. The BET surface area was determined as 741 m$^2$ g$^{-1}$ and an average pore size of about 1.23 nm was confirmed (Figure 1c). Furthermore, the resultant framework showed thermal stability up to 300 °C by thermogravimetric analysis (TGA) (Figure S4). The presence of tightly bound water molecules in the structure was indicated by a weight loss of about 7 % in the range of 100 - 150 °C, which revealed three crystalline water molecules per coordination unit. In agreement with the results from the ICP and EA analysis, a chemical formula of Cu$_2$(C$_{30}$H$_8$O$_8$)(H$_2$O)$_3$ can be determined. In addition, electron paramagnetic resonance (EPR) spectroscopy analysis shows a broad signal at g = 2.122 that can be attributed to the unpaired electrons of the Cu(II) (d$^9$) ion (Figure S5). In the spectrum, the broad line of the Cu(II) superposes signals corresponding to semiquinone radicals which are expected for charge neutral sheets. Superconducting quantum interference device (SQUID) magneti*c*-susceptibility measurements revealed a linear 1/χ-T dependence characteristic for paramagnetic materials (Figure S6). The curve could be fitted to a Curie-Weiss law with θ = -1.98 K, which is indicative of antiferromagnetic interactions between the neighboring metal nodes.[18] An increase in the effective magnetic moment suggests a high-spin magnetic ordering at low temperatures.

We further investigated the crystalline structure of the obtained MOF sample. Powder X-ray diffraction (PXRD) analysis revealed the formation of a crystalline product with prominent peaks at 5.8°, 8.0°, 11.6°, 12.6° 17.9° and 28.1° (Figure S7) corresponding to the d-spacing of 1.53 nm, 1.10 nm, 0.76 nm, 0.70 nm, 0.50 nm and 0.32 nm, respectively. The highly crystalline nature of $Cu_2(OHPTP)$ is also demonstrated by near-atomic-level high-resolution transmission electron microscopy (HRTEM) down to 1.2 Å. The HRTEM images show that $Cu_2(OHPTP)$ exhibits long-range order across the whole rod crystallite and along different crystallographic directions. Figure 2a shows the out-of-plane HRTEM image, which exhibits an ordered π-π stacking along the z-axis with an interlayer distance of 0.31 nm as concluded from the Fast-Fourier-Transform (FFT) of the diffraction pattern. Figure 2b presents the image of the xy-plane after exfoliation of the rod crystals into nanosheets, which shows an extended rhombic network and pores (dark contrast) with a maximum pore diameter of 1.5 nm. Fourier transformed electron diffractions reveal $d$ spacings of 1.58 nm and 1.11 nm which fit well to those deduced from PXRD.

To determine the exact structure of $Cu_2(OHPTP)$ single crystals at the atomic level, we utilized cRED.[19] The cRED data showed unit cell parameters of a=22.74 Å, b=21.58 Å, c=6.50 Å, α=89.33°, β=89.10°, and γ=90.53°, suggesting an orthorhombic unit cell (Figure S9). The 3D reciprocal lattice and 2D planes indicated the following reflection conditions, $hkl$: $h + k = 2n$; $0kl$: $k = 2n$; $h0l$: $h, l = 2n$; and $hk0$: $h + k = 2n$ (Figure 2c). The structure was therefore solved in the space group Cmcm (SI for detailed procedure). Pawley fitting of the structure against the experimental PXRD converged to cell parameters of a = 22.598(2) Å, b = 21.581(3), and c = 6.379(4) Å ($R_P$ = 4.419, $R_{WP}$ = 5.891, $R_{exp}$ = 2.710, GOF = 2.158) and offered an excellent fit with the observed PXRD (Figure 2f and S9). The observed diffractions could be indexed as 110, 200, 202, 220, 310, 130 and 002 planes for the diffractions at 5.8°, 8.0°, 11.6°, 12.6° and 28.1° reflections respectively. The solved structure revealed that Cu(II) cations possess square-planar coordination, forming square-planar [CuO$_4$] units with two OHPTP ligands. Each OHPTP ligand is further surrounded by four [CuO$_4$] units to form a rhombic 2D network with rhombic pores

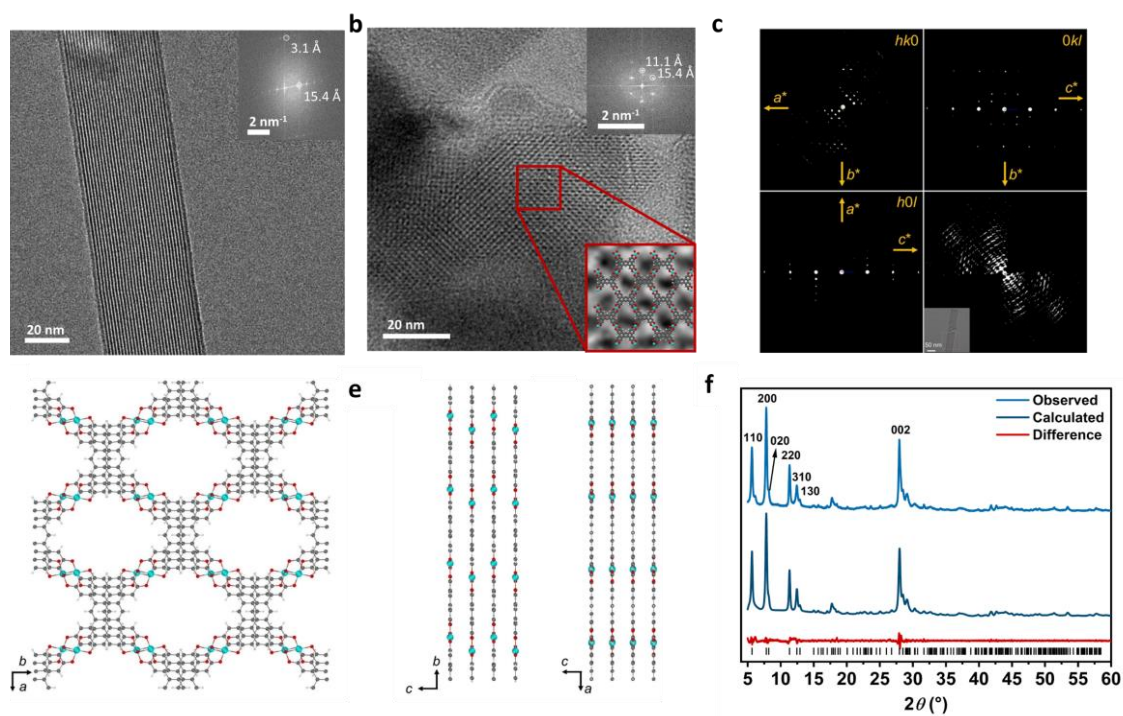

**Figure 2.** (a) Out-of-plane and (b) in-plane high-resolution transmission electron microscopy images with Fourier transformed electron diffractions and structure overlay. (c) Reconstructed 3D reciprocal lattice and 2D slide cuts from cRED data. (d) In-plane structural model; (e) out-of-plane structure models along different crystallographic directions. (f) Experimental and Pawley-refined crystal PXRD showing a good fit for the peaks corresponding to the 110, 200, 020, 220, 310 and 002 reflections.

(Figure 2 d). $Cu_2(OHPTP)$ possesses a layer stacked structure with a short interlayer distance of 0.32 nm. Notably, the 2D layers adopt an alternating stacking order where two layers are shifted only along the b-axis (Figure 2e), leading to an AB stacking with an offset close to 2 Å. The stacking results in an overlap of the aromatic cores while preventing a direct stacking of Cu atoms. The position of the metal nodes leads to a distorted pseudo-octahedral coordination environment for the Cu(II) with the $CuO_4$ unit in the equatorial position with bond lengths of 2.0 and 1.9 Å and the axial oxygen atoms originating from the adjacent layers with 3.2 Å distance.

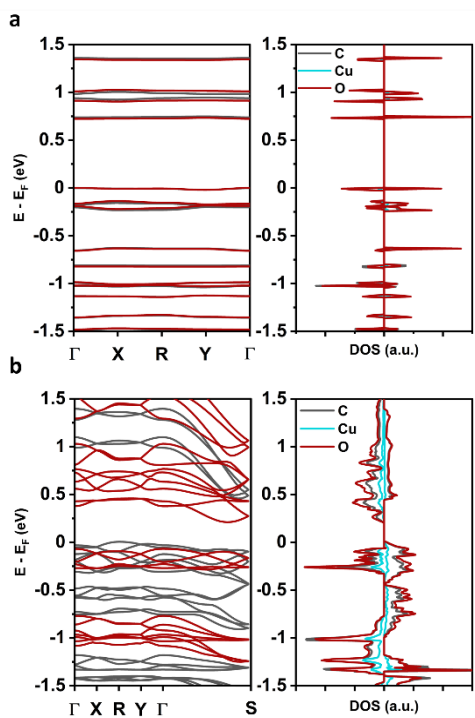

**Figure 3.** Band structure and density of states (DOS) projected on atoms of $Cu_2(OHPTP)$. (a) Monolayer and (b) layered bulk model.

The band structures of monolayer and bulk $Cu_2(OHPTP)$ were calculated using DFT on the resolved crystal structure. After initial geometry optimization (relaxation of atomic positions; SI for detailed parameters), the band structure and electronic density-of-states (DOS) were calculated at the DFT/HSE06 level of theory with the POB-TZVP basis set as implemented in Crystal17. The monolayer adapts a spin-paired state, and the band structure shows nearly flat bands with a band gap of 0.75 eV. The DOS in the energy window of +/- 1.5 eV around the Fermi level is mostly localized at the ligand or metal atoms, indicating a limited π-conjugation within the plane (Figure 3a). In the layer-stacked case, the energetically most favorable spin state is a high-spin state with 16 unpaired electrons per unit-cell, which also corresponds to the results of SQUID measurements. The unpaired spins are mainly localized at the Cu and O centers in the linkage. The multilayered MOF is an indirect bandgap semiconductor with a bandgap of around 0.61 eV and 0.45 eV for alpha and beta spin channels, respectively (Figure 3b). Contrary to the monolayer, the band dispersion is more prominent and rather high in the out-of-plane direction. However, visible dispersion appears also in-plane upon stacking. The effective masses of the electron and holes in the VB and CB were calculated as $m_e = 0.78\ m_0$ and $m_h = 1.21\ m_0$ and the harmonic mean as $m^* = 0.96\ m_0$. The electronic states show high conjugation and hybridization between Cu, C and O atoms. As suggested from the calculations, the π-d orbital overlap leading to the electron delocalization is only visible in the bulk system supporting the importance of the stacking interactions and proximity effects on the overall transport properties previously calculated theoretically.[20]

Solid state UV/Vis – near IR absorption spectroscopy of $Cu_2(OHPTP)$ coated on quartz glass shows two strong absorption bands in the visible region at 320 and 620 nm, respectively, and one broad and weak band in the NIR range at 1350 nm (Figure 4a). The first two can be ascribed to π-π* and metal-to-ligand charge-transfer bands,[9] the absorption in the NIR range is characteristic of highly-conjugated, polymeric materials.[21] Due to the low intensity, the optical energy gap was determined from the UV/Vis – near IR absorption spectrum via the Tauc method for an indirect transition (Figure S10). The determined indirect gap is ~0.5 eV, which is comparable to the DFT calculation. Ultraviolet-photoemission spectroscopy (UPS) was used to probe the energy of the valence band level with respect to $E_F$ (Figure 4b, Figure S11). No clear valence band maximum appeared under a bias of 5 V and a constant decrease of electron density up to $E_F$ was observed indicating the presence of electronic states close to the Fermi-level, suggesting strong *p*-type doping in the material.

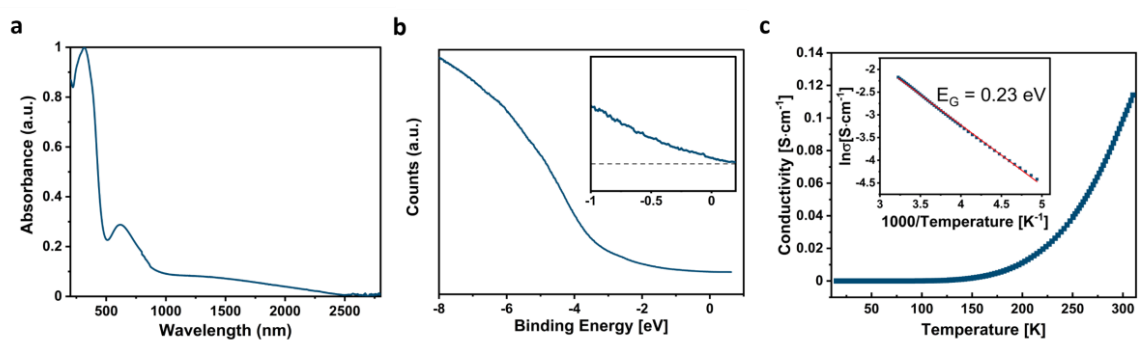

**Figure 4.** (a) UVvis-NIR spectrum of $Cu_2(OHPTP)$ on quartz glass. (b) UPS of a thin layer of $Cu_2(OHPTP)$ on gold coated $SiO_2$. (c) 4-probe van-der-Pauw variable temperature-conductivity measurements of a sample pellet with Arrhenius-plot.

To assess the electrical properties of the material, the conductivity of pellets made from MOF powder was determined via a four-probe measurement in the van-der-Pauw (vdP) geometry under argon. After Ohmic contacts were established, $Cu_2(OHPTP)$ showed a nonlinear increase of conductivity with temperature (Figure 4c, Figure S12), indicating a typical semiconducting behavior where charge carrier transport is thermally activated. A conductivity of 0.1 S cm$^{-1}$ at 300 K could be determined. The high-temperature region of the conductivity-temperature plot could be fitted by an Arrhenius-type function, providing a thermal energy gap of ~0.23 eV. In the temperature range between 40 to 300 K, the conductivity data could be properly modelled by a 3D Mott variable range hopping model. This indicates that charge carrier hopping dominates the charge transport between different single crystals which is often observed in polycrystalline samples.

In order to investigate the charge carrier mobility and doping contributions to the DC conductivity we performed terahertz time-domain spectroscopy (THz-TDS) on polycrystalline samples of Cu$_2$(OHPTP). This technique probes the frequency-resolved complex conductivity of the sample in a non-contact fashion.[22] To access this information, a regular measurement compares the changes in amplitude and phase in transmission for a freely propagating ~1THz pulse through the air and the sample (SI for detailed conditions). The THz-TDS results on a 280 µm thick powder sample sandwiched between fused silica windows at 300K are presented in Figure 5a. The conductivity of the sample can be properly described by the Drude-Smith (DS) model (see SI). A phenomenological model that applies to free carriers experiencing some degree of localization induced by backscattering, which in the DS model is weighted by the backscattering parameter $c_{DS}$ (see SI).[23] The latter can take values between 0 and -1, representing a lack or total localization of free charge carriers within the probed mean free path, respectively. From the DS model we obtain a scattering time of $\tau_s = 0.205\,ps$ and a plasma frequency of $\omega_p = 13.28\,\text{THz}$. Taking into account the averaged effective mass estimated theoretically ($m^* = 0.96 \cdot m_0$) and the value $c = -0.97$ obtained from the DS fit, we can obtain a charge carrier mobility ($\mu = \left(\frac{\tau \cdot e}{m^*}\right)[1 + c]$) and carrier density of $\mu = 10.0\,cm^2V^{-1}s^{-1}$ and $N = 6.7 \cdot 10^{16}\,cm^{-3}$ respectively. The value obtained for the mobility is rather large taking into account that the obtained $c_{DS}$ parameter almost equals 1; in this respect, if the DS model is accurate in describing the data, one could extrapolate (for $c_{DS}$=0) the expected value for the mobility in a macroscopic single crystal as $\mu = 300\,cm^2V^{-1}s^{-1}$. Finally, it is worth highlighting here that the DC conductivity of the bulk sample extrapolated from the DS model to the zero-frequency limit ($\sigma_{DC} = 0.1\,S\,cm^{-1}$) matches the value obtained

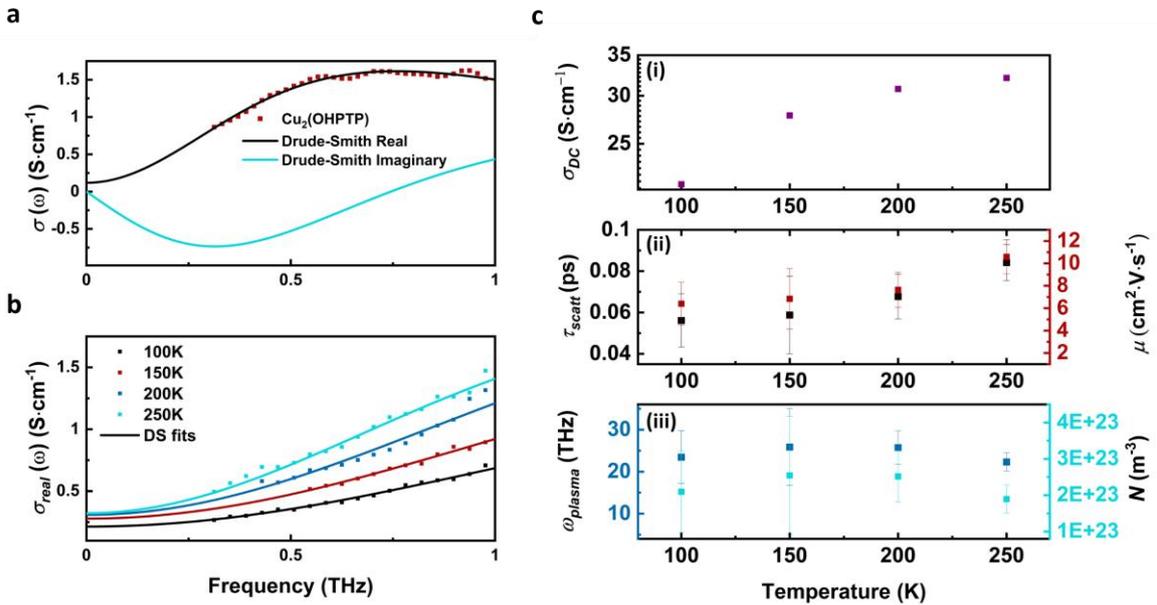

**Figure 4.** (a) Frequency-resolved real conductivity of Cu$_2$(OHPTP) powder (thickness of 280 µm). The red solid squared data points correspond to the experimentally derived values under ambient conditions. The corresponding solid lines are best fit with the Drude-Smith model (see eq. S3), giving access to the real (black solid line) and imaginary (blue solid line) conductivity. (b) Frequency-resolved real conductivity of Cu$_2$(OHPTP) powder (thickness of 150µm) with increasing temperature in vacuum conditions. The solid squared datapoints correspond to the experimentally derived values at 100 K (black), 150 K (red), 200 K (green) and 250 K (blue). The corresponding solid lines are fits to the Drude-Smith model (see eq. S1). (c) (i) Static conductivity ($\sigma_{DC}$, purple squares), (ii) scattering time ($\tau_{scatt}$, black open squares) with the corresponding mobility ($\mu$, solid red squares), calculated using eq. S4 and (iii) plasma frequencies ($\omega_{plasma}$, open green squares) with the corresponding carrier density (N, solid blue squares), calculated using eq. S5. The error bars for $\tau_{scatt}$ and $\omega_{plasma}$,

by using 4-probe contact techniques (Figure 4d); which seems to support the data and DS modelling.

To obtain information about the charge transport mechanism in the bulk sample, we performed temperature-dependent THz-TDS experiments in Cu$_2$(OHPTP) powder (150 µm thick) under vacuum (see SI). From bare data (see Figures 5b and c(i)) we resolved a decay of the DC conductivity, in qualitative agreement with the 4-probe data. From DS modeling we obtain the T variation of the scattering rate and plasma frequency (Figure 5c (ii) and (iii)). While the plasma frequency of the sample and consequently the carrier density remain largely unaffected by increasing temperature, the scattering times (and hence mobilities) increase with increasing temperature (see Figure 5c). This suggests a thermally activated hopping-type charge transport mechanism, consistent with the results obtained on DC probes, a transport mechanism that is often characteristic for polycrystalline samples. Finally, we need to note that the backscattering parameter $c_{DS}$, remains fairly constant for all temperatures (see SI, Figure S13).

# Conclusion

Herein, we demonstrated the synthesis of a new Cu$_2$(OHPTP) semiconducting 2D *c*-MOF single crystal with rhombic lattice, a defined band gap and high intrinsic charge carrier mobility and conductivity. The µm large single-crystals allow the experimental determination of the crystalline structure at the atomic level by imaging and diffraction technics. The Cu$_2$(OHPTP) is a p-type semiconductor with an indirect bandgap of 0.5 eV and exhibits high room-temperature electrical conductivity of 0.1 S cm$^{-1}$. The DFT calculations suggest that the large aromatic core of OHPTP leads to improved π-π interactions along the direction normal to the plane, which contributes to strong band dispersion close to the Fermi level and dominates the overall conductivity. Terahertz spectroscopy analysis demonstrated a high charge mobility of 10 cm$^2$V$^{-1}$s$^{-1}$ which is a record among semiquinoid 2D *c*-MOFs. Our work paves the path to the development of new intrinsically semiconducting MOFs with high charge mobilities for future logic devices.

# Acknowledgements

#L.S., G.Z. and M.W. contributed equally to this work. The authors acknowledge cfaed and Dresden Center for Nano- analysis (DCN) at TUD. This work is financially supported by ERC starting grant (FC2DMOF, No. 852909), ERC Consolidator Grant (T2DCP), DFG projects (CR*C*-1415, No. 417590517), EMPIR-20FUN03-COMET, as well as the German Science Council and Center of Advancing Electronics Dresden (cfaed). R.D. thanks Taishan Scholars Program of Shandong Province (tsqn201909047) and National Natural Science Foundation of China (22272092). E.C. acknowledges MCIN/AEI grant PID2019-107808RA-I00 and Comunidad de Madrid grants 2021-5A/AMB-20942 & P2018/NMT-451. V.B. acknowledges MSCA HORIZON 2020 funding via 101030872 – PhoMOFs grant. Z.H. acknowledges financial support by the Swedish Research Council Formas (2020-00831), and the Swedish Research Council (VR, 2022-02939). We thank Zichao Li and Yunxia Zhou (HZDR Dresden-Rossendorf) for conductivity and squid measurements, Marielle Deconinck (TU Dresden) for UPS measurements, Darius Pohl (DCN) for TEM analysis and Guangbo Chen (TU Dresden) for EPR measurements and Yannan Liu (TU Dresden) for initial structure modeling. We acknowledge the use of the facilities at Dresden Center for Nanoanalysis at Technische Universität Dresden. The authors thank the